\documentclass[preprint,aps,prc,showpacs,nofootinbib]{revtex4}
\usepackage{epsfig}
\usepackage{graphicx,amsmath}

\newcommand\ba{\begin{eqnarray}}
\newcommand\ea{\end{eqnarray}}
\newcommand\be{\begin{equation}}
\newcommand\ee{\end{equation}}
\newcommand\nn{\nonumber}

\newcommand{\br} [1]{ \left( #1 \right) }
\newcommand{\brs}[1]{ \left[ #1 \right] }
\newcommand{\brf}[1]{ \left\{ #1 \right\} }

\begin{document}
\title{Comments to the calculation of transverse beam spin asymmetry for electron proton elastic scattering}
\author{E.~A.~Kuraev}
\affiliation{\it JINR-BLTP, 141980 Dubna, Moscow region, Russian Federation}
\author{S.~Bakmaev}
\affiliation{\it JINR-BLTP, 141980 Dubna, Moscow region, Russian Federation}
\author{V.~V.~Bytev}
\affiliation{\it JINR-BLTP, 141980 Dubna, Moscow region, Russian Federation}
\author{Yu.~M.~Bystritskiy}
\affiliation{\it JINR-BLTP, 141980 Dubna, Moscow region, Russian Federation}

\author{E.~Tomasi-Gustafsson}
\affiliation{\it DAPNIA/SPhN, CEA/Saclay, 91191 Gif-sur-Yvette
Cedex, France }

\date{\today}

\begin{abstract}
The transverse beam spin induced asymmetry is calculated for
the scattering of transversally polarized electrons on a proton target within a  realistic model. Such asymmetry is due to the
interference between the Born amplitude and the imaginary part of two photon exchange amplitude. In particular, the contribution of non-excited hadron state (elastic) to the two photon amplitude is calculated. The elastic contribution requires infrared divergences regularization and can be expressed in terms of  numerical integrals of the target form factor. The inelastic channel
corresponding to the one pion hadronic state contribution is enhanced by squared logarithmic terms. We show that the ratio of elastic over inelastic channel
is of the order of 0.3 and cannot be ignored. Enhancement effects due to the decreasing of
form factors bring the transverse beam asymmetry to values as large as $10^{-4}$ for particular kinematical conditions.
\end{abstract}

\maketitle
\section{Introduction}

Polarization observables are a powerful tool for a precise investigation of the
 nucleon structure. Elastic electron proton scattering is the simplest reaction
which gives information on the dynamical properties of the nucleon, as electromagnetic
form factors (EM FFs), if one assumes one-photon exchange to be the underlying reaction
 mechanism. With the advent of high energy, highly polarized electron beams and hadron
 polarimeters in the GeV range, it has become possible to measure polarization observables
in high precision experiments up to large values of the momentum transfer squared.
In particular, in elastic $ep$ scattering, the polarization of the proton induced by
a longitudinally polarized beam allows to access the proton EM FFs ratio, as firstly
proposed in \cite{Ak68,Ak74}.

The ratio between the longitudinal and the transverse polarization of the proton in
the scattering plane has been measured up to a value of the momentum transfer squared, $Q^2$=$-q^2$=$-t$=5.8 GeV$^2$, giving surprising
results \cite{Jo00}. It has been found that the electric and magnetic distribution
in the proton are different, and that the electric FF decreases faster with $Q^2$ than
previously assumed. However, measurements based on the Rosenbluth separation
(unpolarized $ep$ elastic cross section) give contradicting results, and confirm that
the ratio of electromagnetic proton FFs is compatible with unity.

The extraction of FFs by polarized and unpolarized experiments is based, in both cases, on the
same formalism, which assumes one photon exchange. To explain the discrepancy between
these results, it has been suggested that two photon exchange should be taken into
account \cite{Gu03}. Although such mechanism is suppressed by a factor of
$\alpha$,  due to the
steep decreasing of the electromagnetic FFs with $Q^2$, two-photon exchange
 where the momentum transfer is shared between the two photons, can become
important with increasing $Q^2$. This fact was already indicated in the seventies
\cite{Gu73}, but it was never experimentally observed. Recently, a reanalysis of the
 existing data in deuteron \cite{Re99} and in proton \cite{Etg05} did not show
evidence for such mechanism (more precisely the real part of the interference between
 $1\gamma$ and $2\gamma$ exchange), in the last case at the level of 1\%. Although there are a few
calculations which show that the box diagram contribution is too small to solve the FF
ratio discrepancy \cite{Bo06,By06}, it is interesting to study the implications of the
two photon mechanism as one expects a detectable contribution of two-photon exchange in
 electron hadron scattering, with increasing $Q^2$.
The relative role of the two-photon exchange with respect to the main (one photon)
contribution is expected to be even larger for $d$ or $^3\!He$, due to the steep
decreasing of the electromagnetic FFs. The presence of the two photon exchange amplitude should be firstly unambiguously experimentally detected. It would induce
an additional structure function and more complicated expressions for all the
observables. An exact calculation of two photon exchange in
frame of QED for $e \mu$ elastic scattering was done in \cite{target},
where it was found that its contribution to charge asymmetry is very small (at percent level). This statement must be valid also for $ep$ scattering, as $e\mu$ scattering
can be considered to give an upper limit for the proton case.

Other observables, which vanish in the one-photon
approximation, can be more sensitive to the presence of this mechanism. For example, the transverse beam spin asymmetry (TBSA)
for the case of scattering of transversally polarized electrons on a proton target
\be
e(\vec a,p_1)+p(p) \to e(p_1')+p(p'),
\label{Eq:eq1}
\ee
(the momenta of the particles are shown in brackets and $\vec a$ is the electron polarization)
 is defined as
\be
|\vec a_\perp|A(s,t)=\displaystyle\frac{
\displaystyle\frac{d\sigma}{d\Omega}(\vec a)-
\displaystyle\frac{d\sigma}{d\Omega}(-\vec a)
}{
\displaystyle\frac{d\sigma}{d\Omega}(\vec a)+
\displaystyle\frac{d\sigma}{d\Omega}(-\vec a)
},\label{Eq:eq2}
\ee
where $d\Omega $ is the phase space volume of the scattered electron,  $|\vec a_\perp|$
 is the component of electron spin, which is transversal to the electron scattering plane, and $A(s,t)$ is
the analyzing power. TBSA is function of two Mandelstam variables, the total energy
$s=(p_1+p)^2-M^2=2pp_1$ and the momentum transfer  $t=(p_1-p_1')^2=-2p_1p_1'$.
Such polarization observable was recently measured in high precision experiments
devoted to parity violation in elastic $ep$ scattering, which have a sensitivity
of 10$^{-6}$ i.e., particle per million (ppm). Indeed, nonzero values for this
asymmetry were found: $A=15.4 \pm 5.4$ ppm at $Q^2=0.1$ (GeV/c)$^2$ in the SAMPLE
experiment, at large scattering angle \cite{We01} whereas in the MAMI experiment
 $A=(-8.59 \pm 0.89_{stat} \pm 0.75_{syst})$ ppm at $Q^2$= 0.106 (GeV/c)$^2$, and
 $A=(-8.52 \pm 2.31_{stat} \pm 0.87_{syst})$ ppm at $Q^2$= 0.230 (GeV/c)$^2$ \cite{Ma05}.

Such observable is sensitive to the imaginary part of the two-photon amplitude,
more exactly to the interference of Born and imaginary part of box amplitude and vanishes
in the Born approximation.

TBSA was recently calculated in in a series of
papers. In Ref. \cite{Gorch}
a general analysis of the double Compton amplitude was derived in terms of 18 kinematical
amplitudes. The kinematics was restricted to the case when the two virtual photons have
the same virtuality. The results for TBSA at forward angle, in the energy range 3$\div$ 45 GeV
are negative, of the order of few ppm, and are not in contradiction with the experimental data.

In Ref.  \cite{Mer} attention was paid to inelastic hadron states, and the
spin asymmetry was shown to be enhanced by double logarithmic terms
$\sim \ln^2(-t/(m_e^2))$ for small $t$ kinematics. It was stated that the the
contribution  to TBSA from the proton intermediate
state is suppressed by several orders of magnitude. The excitation of $\Delta(1232)$ resonance
intermediate state was also considered. Indeed, $\Delta(1232)$  and any nucleon resonance
can not exist in the intermediate state as real particles, and a virtual or off-mass shell
$\Delta$ resonance has no physical meaning. In our opinion, it is more realistic to consider, instead, a definite mechanism, with production of a $\pi+N$ intermediate state, which is the simplest inelastic channel.

Inelastic contributions can be expressed in terms of the total photoproduction cross section. This assumption is however an approximation because the exchanged photons are off mass shell and $t\neq 0$ \cite{Mer}.

In the intermediate hadronic state, elastic as well as inelastic contributions should be taken into account. In lowest order of perturbation theory (PT) the elastic intermediate state contribution suffers from infrared divergences, but contributes to TBSA. When considering
up down proton spin asymmetry, this kind of contribution was not properly considered in the previous literature, starting from the basic references \cite{Ruj}, as
well as in more recent works. In the present work, the contribution of elastic hadronic state is taken into account in terms
of a general eikonal phase, and an IR cancellation procedure (IR regularization) is applied.
After this procedure, it remains a finite contribution to TBSA, which, at our knowledge, was not
calculated in previous works.

The purpose of this paper is to point out shortcomings of the previous work, to drive the
attention to the necessity of considering correctly the elastic intermediate state  and to compare its relative contribution to the simplest inelastic process of pion production.

The present calculation differs from the published ones mainly in two respects. We show that after applying the
infrared regularization procedure, the contribution of elastic intermediate states
is not suppressed as it was stated in \cite{Afkar}, it produces a finite contribution in electron mass zero limit
and can be expressed in terms of target FFs. A definite channel of pion production in the intermediate state is calculated, in order to
estimate the inelastic hadronic states.  Off-mass shell photons and kinematics with $t\ne 0$ are considered. For the numerical estimation  we take a simple ansatz for nucleon FFs, (dipole approximation for the Pauli FF  and vanishing Dirac FF) which is justified, in principle, only at small or at very large values of $Q^2$. Nevertheless, it allows to carry on a simple formalism, and does not affect qualitatively the results. The present calculation can be generalized to realistic FFs in a straightforward way.

\section{Formalism}

\begin{figure}
\begin{center}
\includegraphics[bb=0 0 1911 503, scale=0.22]{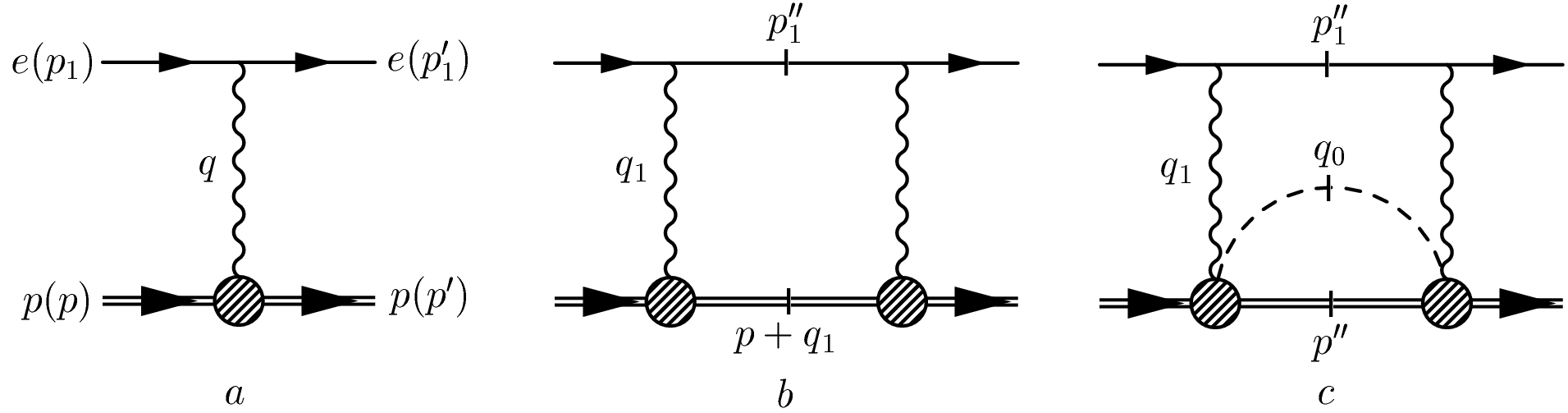}
\caption{Polarized electron-proton elastic scattering:
$a$ - Born approximation, $b$, $c$ - imaginary part of two photon
exchange amplitude ($b$ - elastic intermediate state, $c$ - inelastic
intermediate state).}
\label{Fig:Diagrams}
\end{center}
\end{figure}

The amplitude of elastic electron-proton-scattering, Eq. (\ref{Eq:eq1}), taking into account
only exchanged virtual photons, can be written as \cite{Bethe}:
\ba
M(s,t)=e^{i\phi(t)} M_1(s,t),
\quad \phi(t)=\frac{\alpha}{\pi}c\ln\left(\frac{-t}{\lambda^2}\right),
\label{Eq:eq3}
\ea
where $\lambda$ is a fictitious photon mass and $M_1$ is finite at the limit $\lambda \to 0$,
$c$ is a real constant, depending on FFs and it is not calculated here.

In the language of Feynman diagrams, the phase factor $e^{i\phi}$ can be associated
with single hadron intermediate hadronic states, whereas $M_1$ corresponds to inelastic
intermediate hadronic states such as production of neutral pions.

The specific structure of the amplitude, Eq. \ref{Eq:eq3}, suggests a unique way to
calculate TBSA, eliminating infrared divergences. The asymmetry indeed arises from
the interference of the Born amplitude with the imaginary part of the two (or more)
photon exchange amplitude in the lowest order of perturbation theory. The usual
mechanism of infrared (IR) singularities cancellation (taking into account real soft
photon emission) does not work here, since the amplitude of soft photon emission is
real.

The interference of the Born amplitude with the two photon exchange amplitude  where the intermediate state is a proton, leads to a result which can be expressed in the form
$A\phi(t)+B$. Keeping in mind the general form of this scattering amplitude, Eq. \ref{Eq:eq3},
in order to eliminate the terms containing the phase $\phi(t)$, a regularization procedure,
called IR regularization, has to be applied. It consists in dropping terms
which contain $\ln(\lambda/m^2)$. This is justified by the cancellation of infrared divergent terms when calculating the matrix element squared (\ref{Eq:eq3}).

The general form for the contribution to the differential cross section for the case of transversally
polarized incident electron is related to the factor
\ba
m(pp_1p'_1a)=m\varepsilon_{\alpha\beta\gamma\delta}p^\alpha p_1^\beta p^{'\gamma}_1
a^\delta\simeq mMEE'\sin\theta|a_\perp|,
\ea
where $ m$, $M$, $E$, $E'=E/\left [1+\frac{E}{M}(1-\cos\theta)\right ]$ are
the mass of electron, the mass of the proton, the energies of the initial and final
electron, respectively,  $\theta$ is the final electron
scattering angle in the laboratory frame, and $a_\perp $ is the degree of transverse polarization of the
incident electron.

The asymmetry is due to the interference of the Born amplitude with the
$s$-channel imaginary part of two-photon exchange amplitude.

\subsection{Elastic contribution}
The elastic contribution to the asymmetry can be written as:
\ba
A_{el}=\frac{2\sum_{spin}M_b M^*_{el}}{\sum_{spin}|M_b|^2|a_\bot|},
\label{Eq:eq5}
\ea
with
\ba
\sum_{spin}|M_b|^2&=&\left(\frac{4\pi\alpha}{t}\right)^28(s^2+u^2+2M^2t)F^2(t), \nn\\
2\sum_{spin}M_bM^*_{el}&=&\frac{2^5\pi\alpha^3}{t}F(t)\frac{s}{s+M^2}\int S_eS_{el}
\frac{dO''}{q^2_1q^2_2}F(t_1)F(t_2),
\label{Eq:eq6}
\ea
where $q_{1,2}^2 = -2\epsilon^2(1-c_{1,2})$ and $\epsilon=s/(2\sqrt{s+M^2})$ in the elastic case. The integration is performed in the center of mass system (CMS) of the initial particles, and  $\epsilon$ is the energy of the electron in this reference frame.
The proton vertex is described by such model for the Pauli and Dirac FFs: $F_1(t)=F(t)=(1+Q^2/Q_0^2)^{-2}$  follows a dipole distribution with $Q_0^2=0.71$ GeV$^2$ and $F_2(t)=0$. This is a good approximation at very large or very small $t$ values. In our case, it is a simple prescription, which allows to factorize the terms containing FFs. The traces $S_e$ and $S_{el}$ have the form
\ba
S_e&=&\frac{1}{4}Tr(\hat p'_1+m)\gamma_\mu(\hat p''_1+m)\gamma_\nu(\hat p_1+m)
\gamma_5\hat a\gamma_\lambda,\\
S_{el}&=&\frac{1}{4}Tr(\hat p'+M)\gamma_\mu(\hat p+\hat q_1+M)\gamma_\nu(\hat p+M)\gamma_\lambda.
\label{Eq:eq7}
\ea
After performing the angular integration (see Appendix \ref{App_inelastic}) we obtain:
\ba
A_{el}&=&\frac{-st\alpha mMEE'\sin\theta}{4(s+M^2)(s^2+u^2+2M^2t) F(t)}Q_{el},
\label{Ael}\\
Q_{el}&=& -12 a_g + (4 s+6 t) a_{1s} + 2(s+t) a_{11} +  [8 M^2 - 2(2 s + t)] a_v - 4M^2 I,
\label{Qel}
\ea
where the integrals $I$, $a_v$, $a_{11}$, $a_{1s}$, $a_g$ are given in the Appendix \ref{App_elastic} (see Eq. (\ref{ElasticIntegrals})).

\subsection{Inelastic contribution}
The contribution of the inelastic channel can be written as:
\ba
A_{inel}&=&\frac{\alpha \cdot t}{64\pi^2}g^2\displaystyle \frac{|F(t)|^{-1}}{s^2+u^2+2M^2t}
\int\frac{\epsilon''_1d\epsilon''_1}{q^2_1q^2_2}S_{e}S_{inel}\frac{dO_\pi dO''_1(s^2-2
\sqrt{s+m^2}\epsilon''_1)}{\pi^2[\sqrt{s-M^2}-\epsilon''_1(1-c_\pi)]^2},\label{Eq:eq11}\\
&&c_\pi=\cos\vec p_1{\! '}\vec q_0, \qquad
 m_e<\epsilon''_1<\frac{s}{2\sqrt{s+M^2}}\left(1-\frac{m_\pi M^2}{s\sqrt{s+M^2}}\right),
\nn
\ea
where the value of $g^2\le 3$ can be extracted from the total photon-proton cross section
$\sigma_{tot}^{\gamma p}\approx 0.1$ mb $> \alpha g^2/(4m_\pi^2)$. The trace has the form:
\ba
S_{inel}&=&\frac{1}{4}Tr(\hat p'+M)\left[\gamma_5\frac{\hat p'-\hat q_0+M}{(p'-q_0)^2-M^2}\gamma_\mu
+\gamma_\mu\frac{\hat p''+\hat q_0+M}{(p''+q_0)^2-M^2}\gamma_5\right]\times \nn \\
&\times&(\hat p''+M)\left[\gamma_\nu
\frac{\hat p-\hat q_0+M}{(p-q_0)^2-M^2}\gamma_5+\gamma_5\frac{\hat p''+\hat q_0+M}{(p''+q_0)^2-M^2}\gamma_\nu\right]
(\hat p+M)\gamma_\lambda ,
\ea
where, in the inelastic case:
\ba
q^2_{1,2}=-\epsilon\epsilon''_1(1-\beta c_{1,2}), \quad 1-\beta^2=
\left[\frac{m_e(\epsilon-\epsilon''_1)}{\epsilon\epsilon''_1}\right]^2.
\ea
In the approximation of quadratic logarithms which is valid when  $\epsilon''_1 \ll M$
the resulting expression for asymmetry can be factorized (in CMS):
\ba
A_{inel}^{as}=-\frac{g^2\alpha mMEE's\sin\theta}{64\pi^2(s^2+u^2+2M^2t)(s+M^2)F(t)} R P,
\label{Ainel}
\ea
where $R$ is the contribution of the nucleon-pion block:
\ba
R &=&
\eta -\frac{-2 M^2+s+2 t}{s} \delta
+\left(\frac{2 M^2}{s}-1\right)\gamma + 2 I_L + \left(4 M^2-t\right) \alpha_v + \nn\\
&&+2 \left(2 M^2+s\right) \beta_v
+\frac{s \left(4 \tau  M^2+t\right)}{4 M^2 (\tau +1)} I_1 + \frac{1}{s} j,
\qquad
\tau=\frac{-t}{4M^2}.
\label{R}
\ea
The coefficients $\eta$, $\delta$, $\gamma$, $I_L$, $\alpha_v$,
$\beta_v$, $I_1$, $j$ are given in Appendix \ref{App_inelastic}, see Eq. (\ref{InelasticIntegrals}),
and the term $P$ is the contribution
of the $2\gamma$-intermediate electron block of amplitude:
\ba
\int\frac{dO''\epsilon_1''d\epsilon_1''}{\pi q_1^2q_2^2}=
-\frac{1}{t}\left[\ln\frac{-t}{m_e^2}\ln\frac{\epsilon^2}{m_e^2}-
\frac{1}{2}\ln^2\frac{\epsilon^2}{m_e^2}+\frac{2\pi^2}{3}\right]=-\frac{1}{t}P.
\label{Pdef}
\ea

\section{Results}

Numerical results are presented in Tables \ref{TableRatio} and \ref{TableTotalAsymmetry} for different values of the outgoing electron scattering angle and of the energy of the initial electron.

The ratio $A_{el}/A^{as}_{inel}$ is given in Table \ref{TableRatio} (see Eqs.  (\ref{Ael},\ref{Ainel}).
The elastic contribution is not negligible, representing more than 30 \% of the inelastic one, for most of the
kinematical region. At small energy and angle it even exceeds the inelastic contribution. The reason is that, at small
energies, inelastic channels are suppressed in vicinity of the pion pair production threshold.

The total asymmetry $A=A_{el}+A^{as}_{inel}$ (in units ppm) is given in \ref{TableTotalAsymmetry}.
It increases when the energy increases, around $\theta=90^0$ where it can be as large as 10$^4$.

The suppression in the vicinity of $\theta=0$ or $\pi$, is driven by the sine term.

In order to compare the calculation with experimental data, different kinematical conditions should
be considered, at low $Q^2$. The experimental values or Ref. \cite{Ma05} are well
reproduced by the present calculation, taking the following values for the cosine of
electron scattering angle: $a=0.64$ and for the pion coupling constant: $g^2=1.5$.


\section{Conclusion}

We calculated the contribution to the electron beam asymmetry in elastic $ep$
scattering, which arises from the two photon exchange amplitude in case
 of non excited hadron intermediate state. The present calculation shows
 that that the elastic contribution can not be neglected, and that it can
 be expressed in terms of hadron FFs.

The present results contradict previous statements of a specific
suppression of this kind of contribution
\cite{Mer,Ruj,Pasq,AAM}, the reason being that this source was
not properly considered. The elastic intermediate state amplitude
suffers from infrared divergences (see Eqs. (1), (3)). This fact
was overlooked in the previous literature.

Inelastic hadronic intermediate states are enhanced in
comparison with the elastic state considered above, and can be expressed
in terms of double virtual
Compton scattering (DVCS) amplitude \cite{Gorch} contrary to statement
\cite{AAM}, where it was expressed in terms of photoproduction cross section.

Double logarithmic enhancement (DL) of inelastic channel takes place at higher
orders of the QED coupling constant, what can lead to Sudakov form-factor suppresion type.
Rigorous investigation of higher orders of QED contributions will be subject of
further investigations.

It must be noted that the results are sensitive to the behavior
of nucleon FFs, but do not chage the qualitatively results of the paper.
 The dipole like decreasing with the
momentum transfer squared  plays a crucial role and results
in a general enhancement of elastic and inelastic channels contributions.

The choice of FFs taken in the present work, allows a simplified calculation.
This approach is realistic at small $Q^2$ and at $Q^2\gg 1$ GeV$^2$
due to the suppression of the $F_2$ form factor. At small $Q^2$, the accuracy of our calculation, due to this approximation is of the order of $\tau$. 

Another approximation, which affects the precision of the present results, is related to the fact that single-logarithmic terms of the form
$\ln(-t/m^2)$  as well as terms of the order $(m_\pi/M)^2$ were omitted. 

Overall, the precision of the present results due to these approximations can be estimated at the level of 10~\%.

\section{Acknowledgments}
One
of us (VB) is grateful to the grant MK-2952.2006.2.
E.A.K. acknowledges the hospitality of DAPNIA/SPhN, Saclay, where part of this work was done.

\begin{table}
\begin{center}
\begin{tabular}{|c|c|c|c|c|c|}
\hline
$E$, GeV &$\theta=30^0$&$\theta=60^0$&$\theta=90^0$&$\theta=120^0$&$\theta=150^0$ \\
\hline
1.0 & -1.377 & 0.138 & 0.292 & 0.331 & 0.340\\
\hline
1.5 & -0.252 & 0.323 & 0.379 & 0.392 & 0.392\\
\hline
2.0 & 0.123 & 0.377 & 0.389 & 0.394 & 0.401\\
\hline
2.5 & 0.261 & 0.375 & 0.378 & 0.381 & 0.377\\
\hline
3.0 & 0.327 & 0.358 & 0.363 & 0.359 & 0.354\\
\hline
3.5 & 0.349 & 0.361 & 0.339 & 0.338 & 0.338\\
\hline
4.0 & 0.348 & 0.335 & 0.327 & 0.314 & 0.316\\
\hline
\end{tabular}
\end{center}
\caption{Ratio of elastic over inelastic terms $A_{el}/A_{inel}^{as}$.}
\label{TableRatio}
\end{table}

\begin{table}
\begin{center}
\begin{tabular}{|c|c|c|c|c|c|}
\hline
$E$, GeV &$\theta=30^0$&$\theta=60^0$&$\theta=90^0$&$\theta=120^0$&$\theta=150^0$ \\
\hline
1.00 & 0.76 & -11.70 & -28.27 & -34.39 & -23.037\\
\hline
1.50 & -2.03 & -22.13 & -43.24 & -43.96 & -26.26\\
\hline
2.00 & -4.47 & -35.28 & -58.57 & -53.37 & -29.87\\
\hline
2.50 & -7.42 & -50.05 & -73.77 & -62.71 & -33.39\\
\hline
3.00 & -11.119 & -66.95 & -88.68 & -71.13 & -37.69\\
\hline
3.50 & -15.73 & -85.08 & -104.42 & -79.98 & -41.58\\
\hline
4.00 & -21.58 & -103.16 & -119.07 & -88.17 & -45.32\\
\hline
\end{tabular}
\end{center}
\caption{Total asymmetry  $A=A_{el}+A^{as}_{inel}$ in units of ppm.}
\label{TableTotalAsymmetry}
\end{table}

\appendix
\section{Elastic contribution (method of integration)}
\label{App_elastic}
The biaxial reference frame is used to parametrize the phase volume $dO''$ with
fixed directions of initial ($\vec{p_1}$) and scattered
($\vec{p_1}{\,\!'}$) electrons.

The angular phase space volume of the intermediate real electron is parametrized as
\ba
\int
dO''=2\int\limits_{-1}^{1}dc_1\int\limits_{c_-}^{c_+}\frac{dc_2}{\sqrt{D}}, \nn
\ea
where
\ba
D&=&1-a^2-c_1^2-c_2^2+2c_1c_2a=(c_2-c_-)(c_+-c_2)>0, \nn\\
c_\pm &=& a c_1 \pm \sqrt{(1-c_1^2)(1-a^2)}, \nn
\label{eq:eqc}
\ea
and
$c_1=\cos\theta_1$, $c_2=\cos\theta_2$, $a=\cos\theta_c$, and $\theta_1=\widehat  {\vec{p}_1 \vec{p_1}{\,\!''}}$,
$\theta_2 =\widehat{\vec{p_1}{\!'} \vec{p_1}{\!''}}$, and  $\theta_c=\widehat {\vec{p}_1\vec{p_1}{\,\!'}}$
(in CMS, where $\vec p_1+\vec p=0$).
Using Euler substitution ($t^2=(c_2-c_-)/(c_+-c_2)$) the relevant integrals are \cite{EK}:
\ba
&&\int^{c_+}_{c_-}\frac{dc_2}{\sqrt{D}}=
\pi;
\quad \int^{c_+}_{c_-}\frac{dc_2}{\sqrt{D}(1-\beta c_2)}=
\frac{\pi}{[(a-\beta c_1)^2+(1-\beta^2)(1-c^2_1)]^\frac{1}{2}};\nn\\
&&
\int^{c_+}_{c_-}\frac{c_2dc_2}{\sqrt{D}}=\pi ac_1; \nn \\
&&
\int\frac{d^2c}{\sqrt{D}(1-\beta_1c_1)(1-\beta_2 c_2)}=\frac{\pi}{\sqrt{d}}\left[\ln\left(\frac{4}{1-\beta^2_1}\right)
+\ln\left(\frac{4}{1-\beta^2_2}\right)+2\ln\frac{1-\beta_1\beta_2a+\sqrt{d}}{4}\right],\nn \\
&& d=(1-\beta_1 \beta_2a)^2-(1-\beta^2_1)(1-\beta^2_2)\nn,
\ea
where $\beta_1$ and $\beta_2$ are  real number and $0<\beta_{1,2}<1$ .

For the case of a proton in the intermediate state one can write the following kinematical relations in CMS:
\ba
&&\vec p_1+\vec p=\vec p_1{''}+\vec p{\,''}=\vec p_1{\,\!'}+\vec p{\,'}=0;\nn\\
&&-\lambda^2+q^2_1=(p_1-p''_1)^2=-\lambda^2-2p^2(1-c_1)=-2\epsilon^2(1-\beta_\lambda c_1);\nn\\
&&q^2=t=-2\epsilon^2(1-a),
\label{eq:eq16}
\ea
with
$$1-\beta^2_\lambda=\displaystyle\frac{\lambda^2}{p^2}, p^2\approx\epsilon^2\approx\displaystyle\frac{s^2}{4(s+M^2)}.$$

Then the following integrals can be calculated:
\ba
I_0&=&\int\frac{dO''}{\pi(q^2_1-\lambda^2)(q^2_2-\lambda^2)}=
-\frac{2}{t\epsilon^2}\ln{\left(\frac{-t}{\lambda^2}\right)},\\
I_1&=&\frac{1}{\pi}\int\frac{dO''}{q_1^2-\lambda^2}=-\frac{1}{\epsilon^2}
\left[\ln\left(\frac{-t}{\lambda^2}\right)+\ln\left(\frac{4\epsilon^2}{-t}\right)\right].\nn
\ea
Including a realistic hadron form factor $F(t)$, with normalization $F(0)=1$, after applying the IR regularization procedure, one  obtains:
\ba
&&IR\int\frac{dO''F(q^2_1)F(q^2_2)}{(q^2_1-\lambda^2)(q^2_2-\lambda^2)}
\left\{1, p''_{1\mu}, p''_{1\mu}p''_{1\nu} \right\}= \nn\\
&&\qquad \left\{I, \quad a_v(p_1+p'_1)_\mu, \quad
a_gg_{\mu\nu}+a_{11}(p_{1\mu}p_{1\nu}+p'_{1\mu}p'_{1\nu})+a_{1s}(p_{1\mu}p'_{1\nu}+p_{1\nu}p'_{1\mu})\right\},
\label{eq:eq19}
\ea
where the scalar coefficients introduced in Eq. (\ref{Qel}) have the form:
\ba
I&=&\frac{1}{2\epsilon^4}\left\{\int\frac{d^2c}{\sqrt{D}\pi}\frac{F(q^2_1)-1}{1-c_1}
\frac{F(q^2_2)-1}{1-c_2}+\right. \nn \\
&&+\left.
\int^1_{-1}\frac{dc_2}{|c_2-a|}\left[\frac{F(q^2_2)-1}{1-c_2}-\frac{F(t)-1}{1-a}\right]+
\frac{F(t)-1}{1-a}L
\right\}, \nn \\
a_v&=&\frac{-1}{t\epsilon^2\pi}\int\frac{d^2cF(q^2_2)(F(q^2_1)-1)}{\sqrt{D}(1-c_1)}-
\frac{1}{t\epsilon^2}\left[\int^1_{-1}\frac{dc_2[F(q^2_2)-F(t)]}{|c_2-a|}+F(t)\cdot L\right],
\label{ElasticIntegrals}\\
a_g&=&\frac{1}{4}ta_{1s}=\frac{1}{t\pi}\int\frac{d^2c}{\sqrt{D}}F(q^2_1)F(q^2_2);
\qquad L=\ln\left[\frac{s^2}{-t(s+M^2)}\right ]\nn\\
a_{11}&=&-\frac{1}{\epsilon^2t^2}
\left\{\int\frac{d^2cq^2_2 F(q_2^2)(F(q^2_1)-1)}{\pi\sqrt{D}(1-c_1)}\right.
\left.+\int^1_{-1}\frac{dc_2}{|c_2-a|}(q^2_2F(q^2_2)-t F(t))+t F(t)\cdot L\right\}. \nn
\ea
It is easy to see that for any form of FFs (particularly for dipole ) all divergences are canceled
and the final expressions are free from discontinuities. For our choice of FF we have:
\ba
\frac{F(q^2_{1,2})-1}{1-c_{1,2}}=\frac{-4\epsilon^2[Q^2_0+\epsilon^2(1-c_{1,2})]}{[Q^2_0+2\epsilon^2(1-c_{1,2})]^2},
\quad \frac{F(t)-1}{1-a}=\frac{-4\epsilon^2[Q^2_0+\epsilon^2(1-a)]}{[Q^2_0+2\epsilon^2(1-a)]^2}.
\ea

Let us recall that the following relation holds between $a$, the cosine of scattering electron angle in CMS, and  the initial electron energy, $\epsilon$:  $1-a={-t}/(2\epsilon^2)$ , and that  $q^2_{1,2}=-2\epsilon^2(1-c_{1,2})$.

\section{Inelastic contribution (useful integrals)}
\label{App_inelastic}
In the inelastic case, after calculating the traces, the
Schouten identity is used to express the loop momenta in the numerator via the denominators (we neglect the terms $m_\pi^2/M^2 \approx 0.02$ compared to those of order unity):
\ba
    c &=& \br{p'-q_0}^2-M^2 = -2 p' q_0, \nn\\
    b &=& \br{p-q_0}^2-M^2 = -2 p q_0, \nn\\
    d &=& \br{p''+q_0}^2-M^2 = 2 p'' q_0. \nn
\ea
Finally, one can write the inelastic contribution in terms
of the following integrals, Eq. (\ref{R}):
\ba
    \int d\omega \displaystyle\frac{1}{b} &=& \int d\omega \displaystyle\frac{1}{c} =
    I_1 = -\frac{1}{4\epsilon^2} L, \qquad
    L=\ln\displaystyle\frac{s+M^2}{M^2},\nn\\
    \int d\omega \displaystyle\frac{b}{c} &=& \int d\omega \displaystyle\frac{c}{b} =
    \displaystyle\frac{1-a}{2\tilde{ \beta}} L + a = I_L,
    \nn \\
    \int d\omega \displaystyle\frac{b^2}{c}&=&
    -2\epsilon^2 \beta
    \brf{
        \frac{1}{\tilde{\beta}}
        \brs{\br{1-a}^2 - \displaystyle\frac{1}{2}\br{1-\tilde{\beta}^2}\br{1-a^2}} L
        +
        2 a - \frac{3}{2}a^2
    } = i,
    \nn \\
    \int d\omega b &=& \int d\omega c= -2\epsilon^2 \tilde{\beta} = j,
    \qquad
    \tilde{\beta} = \displaystyle\frac{s}{s+2M^2}, \nn\\
    \int d\omega \displaystyle\frac{c}{b} q_0^\mu &=& (\gamma p + \delta p')^\mu,
    \qquad
    \gamma = \displaystyle\frac{1}{-16 M^4 \tau (1+\tau)}\left[(2M^2-t) i - 2 M^2 j \right],
    \label{InelasticIntegrals} \\
    \int d\omega \displaystyle\frac{b}{c} q_0^\mu &=& (\gamma p' + \delta p)^\mu,
    \qquad
   \delta = \displaystyle\frac{1}{-16 M^4 \tau (1+\tau)}\left[(2M^2-t) j - 2 M^2 i \right], \nn \\
    \int d\omega \displaystyle\frac{1}{c} q_0^\mu &=& (\alpha_v p + \beta_v Q)^\mu,
    \qquad
    \alpha_v = \displaystyle\frac{1}{s^2}\left[ 2Q^2 - s(M^2+Q^2) I_1 \right], \nn \\
    \int d\omega \displaystyle\frac{1}{b} q_0^\mu &=& (\alpha_v p' + \beta_v Q)^\mu,
    \qquad
    \beta_v = -\displaystyle\frac{1}{s^2}\left[ M^2 + Q^2 +2 M^2 s I_1 \right], \nn \\
    \int d\omega q_0^\mu &=& \eta Q^\mu, \qquad
    \eta = \displaystyle\frac{s}{2Q^2}, \qquad Q = p+p_1, \qquad Q^2 = M^2 + s, \nn
\ea
where the following notation holds for the phase volume element:
\ba
    \frac{d^3 q_0 d^3 p''}{2 \epsilon_0 2\epsilon''}
    \delta^4(Q-p''-q_0)
    =
    \frac{s \pi}{2(s+M^2)} d\omega,
    \qquad
    d\omega = \frac{dO_\pi}{4\pi}.
\ea

\end{document}